\documentclass[aps,showpacs,superscriptaddress,
preprintnumbers,twocolumn]{revtex4}%
\usepackage{amsfonts}
\usepackage{amsmath}
\usepackage{hyperref}
\usepackage{scalefnt}
\usepackage{amssymb}
\usepackage{graphicx}%
\newtheorem{theorem}{Theorem}

\newtheorem{proposition}[theorem]{Proposition}

\begin{document}
\title{Two-colorable graph states with maximal Schmidt measure}
\author{Simone Severini}
\email{ss54@york.ac.uk}
\affiliation{University of York, U.K.}

\begin{abstract}
The Schmidt measure was introduced by Eisert and Briegel for quantifying the
degree of entanglement of multipartite quantum systems [\emph{Phys. Rev.
A}\textbf{\ 64}, 022306 (2001)]. Although generally intractable, it turns out
that there is a bound on the Schmidt measure for two-colorable graph states
[\emph{Phys. Rev. A} \textbf{69}, 062311 (2004)]. For these states, the
Schmidt measure is in fact directly related to the number of nonzero
eigenvalues of the adjacency matrix of the associated graph. We remark that
almost all two-colorable graph states have maximal Schmidt measure and we
construct specific examples. These involve perfect trees, line graphs of
trees, cographs, graphs from anti-Hadamard matrices, and unyciclic graphs. We
consider some graph transformations, with the idea of transforming a
two-colorable graph state with maximal Schmidt measure into another one with
the same property. In particular, we consider a transformation introduced by
Fran\c{c}ois Jaeger, line graphs, and switching. By making appeal to a result
of Ehrenfeucht \emph{et al}. [\emph{Discrete Math.} \textbf{278} (2004)], we
point out that local complementation and switching form a transitive group
acting on the set of all graph states of a given dimension.


\end{abstract}
\maketitle

\section{Introduction}

Graph states are certain pure multi-party quantum states associated to graphs
(see, \emph{e.g.}, \cite{rbb, s1, s}). Graph states have important
applications in quantum error correction and in the one-way quantum computer
(see \cite{g} and \cite{b}, respectively). The graph states associated to
bipartite graphs, also called \emph{two-colorable graph states} \cite{a}, have
useful properties. For example, Greenberger-Horne-Zeilinger states and cluster
states are two-colorable graph states \cite{h}. In addition, two-colorable
graph states are equivalent (up to local unitary transformations) to
Calderbank-Shor-Steane states \cite{kl}.

The Schmidt measure, based on a generalization of the Schmidt rank of pure
states, was introduced by Eisert and Briegel \cite{e} for quantifying the
degree of entanglement of multipartite quantum systems. Although generally
intractable, it turns out that it is easy to bound the Schmidt measure for
two-colorable graph states. The Schmidt measure is in fact directly related to
the rank of the graph associated to the state. (It has to be remarked that an
entanglement measure for graph states, which can be computed efficiently from
a set of generators of the stabilizer group, was introduced in \cite{cu}.) The
\emph{rank of a graph} is the number of nonzero eigenvalues of the adjacency
matrix; its study has a number of applications \cite{c}. For example, in
chemistry, the graph representing the carbon-atom skeleton of a molecule has
full rank (that is, equal to the number of vertices) if the so-called
conjugated molecule is chemically stable.

This paper presents a portfolio of two-colorable graph states with
maximal\ Schmidt measure. Practically, this paper mainly surveys results on
$(0,1)$ invertible matrices and merely translates these into the context of
graph states. The paper is organized as follows. In Section 2, we recall the
notion of graph state and Schmidt measure. In Section 3, we describe some
classes of graphs with full rank. These concern perfect trees, certain
bipartite graphs obtained from a class of cographs (called here B\i y\i
ko\u{g}lu cographs), certain graphs obtained from anti-Hadamard matrices, and
unicyclic graphs. In Section 4, we consider some graph transformations, with
the idea of transforming a two-colorable graph state with maximal Schmidt
measure into another one with the same property. In particular, we consider an
operation introduced by Fran\c{c}ois Jaeger, line graphs, and switching. Even
if tt does not seem easy to establish some general relation between switching
and the rank of the adjacency matrix, we claim that switching is worth further
study. In fact, together with local complementation (a graph operation which
is central in the context of graph states), it generates a transitive group
acting on the set of graph states of a given dimension.

\section{Graph states}

Firstly, let us recall the definition of a graph state \cite{s}. A
\emph{graph} $G=(V(G),E(G))$ is a pair whose elements are two set,
$V(G)=\{1,2,...,n\}$ and $E(G)\subset V(G)\times V(G)$. The elements of $V(G)$
and $E(G)$ are called \emph{vertices} and \emph{edges}, respectively. We
assume that $\{i,i\}\notin E(G)$ for all $i\in V(G)$. The \emph{adjacency
matrix} of $G$ is the matrix $A(G)$ such that $A(G)_{i,j}=1$ if $\{i,j\}\in
E(G)$ and $A(G)_{i,j}=0$ if $\{i,j\}\notin E(G)$. The matrices%
\[%
\begin{tabular}
[c]{lllll}%
$\sigma_{x}=\left[
\begin{array}
[c]{cc}%
0 & 1\\
1 & 0
\end{array}
\right]  ,$ &  & $\sigma_{y}=\left[
\begin{array}
[c]{cc}%
0 & -i\\
i & 0
\end{array}
\right]  $ & and & $\sigma_{z}=\left[
\begin{array}
[c]{cc}%
1 & 0\\
0 & -1
\end{array}
\right]  $%
\end{tabular}
\]
are called \emph{Pauli matrices}. Let $I$ be the identity matrix. Given a
graph $G$, we define a block-matrix $S(G)$ with $ij$-th block defined as
follows:%
\[
S(G)_{i,j}=\left\{
\begin{tabular}
[c]{ll}%
$\sigma_{x},$ & if $i=j;$\\
$\sigma_{z},$ & if $\{i,j\}\in E(G);$\\
$I,$ & if $\{i,j\}\notin E(G).$%
\end{tabular}
\right.
\]
From the block-rows of $S(G)$ we construct the following matrices:%
\begin{align*}
S(G)_{1}  &  =S(G)_{1,1}\otimes S(G)_{1,2}\otimes\cdots\otimes S_{1,n},\\
S(G)_{2}  &  =S(G)_{2,1}\otimes S(G)_{2,2}\otimes\cdots\otimes S_{2,n},\\
&  \vdots\\
S(G)_{n}  &  =S(G)_{n,1}\otimes S(G)_{n,2}\otimes\cdots\otimes S_{n,n}.
\end{align*}
It can be shown that these matrices all commute. The \emph{graph state}
associated to the graph $G$ is defined to be the common eigenvector of the
matrices $S(G)_{1},S(G)_{2},...,S(G)_{n}$ with eigenvalue $1$. We denote by
$|G\rangle$ the graph state corresponding to the graph $G$.

Let $\mathcal{H\cong H}_{1}\mathcal{\otimes H}_{2}\mathcal{\otimes
\cdots\otimes H}_{n}$ be an Hilbert space assigned to a quantum system with
$n$ subsystems. The pure state $|\psi\rangle$ of the system can be written as%
\begin{equation}
|\psi\rangle=\sum_{i=1}^{R}\alpha_{i}|\psi_{i}\rangle_{1}\mathcal{\otimes
}|\psi_{i}\rangle_{2}\mathcal{\otimes\cdots\otimes}|\psi_{i}\rangle_{n},
\label{sum}%
\end{equation}
where $\alpha_{i}\in\mathbb{C}$ for $i=1,2,...,R$, and $|\psi_{i}\rangle
_{j}\in\mathcal{H}_{j}$ for $j=1,2,...,n$. The \emph{Schmidt measure} of
$|\psi\rangle$ is defined by $E_{S}(|\psi\rangle)=\log_{2}(r)$, where $r$ is
the minimal number $R$ of terms in the summation of Eq. (\ref{sum}) over all
linear decompositions into product states.

Two vertices $i,j$ of a graph are said to be \emph{adjacent} if $\{i,j\}$ is
an edge (and the edge is then \emph{incident} with the vertices). A graph is
\emph{bipartite} if it has a bipartition of the set of vertices into two
disjoint sets where vertices in one set are adjacent only to vertices in the
other set. The \emph{spectrum} of a graph $G$ is the collection of eigenvalues
of $M(G)$, or equivalently, the collection of zeros of the characteristic
polynomial of $M(G)$ (see, \emph{e.g.}, \cite{c}). The \emph{rank} of $G$,
denoted by $r(G)$, is the number of non-zero eigenvalues in the spectrum of
$M(G)$. The \emph{nullity} of $G$ is the number of eigenvalues of $M(G)$ which
are equal to zero. A number of important parameters of a graph $G$ is bounded
by a function of $r(G)$, for example, clique number, chromatic number,
\emph{etc.} The rank of $G$ is bounded above by the number of distinct nonzero
rows of $M(G)$. A graph is \emph{nonsingular} if $r(G)=|V(G)|$. The following
proposition gives bounds on the Schmidt measure of graph states associated to
bipartite graphs:

\begin{proposition}
[\cite{e}]\label{bou}Let $G$ be a bipartite graph. Then
\[
\frac{1}{2}r(M(G))\leq E_{S}(|G\rangle)\leq\left\lfloor \frac{|V|}%
{2}\right\rfloor .
\]
Moreover, if $r(G)=n$ then $E_{S}(|G\rangle)=\left\lfloor \frac{|V|}%
{2}\right\rfloor $.
\end{proposition}

Before to move on to the next sections, it is useful to describe a way to
construct a bipartite graph on $2n$ vertices from an $n\times n$
$(0,1)$-matrix (recall that a $(0,1)$\emph{-matrix} is a matrix whose entries
are in the set $\{0,1\}$). Let $G$ be a graph with adjacency matrix%
\[
M(G)=P\left(
\begin{array}
[c]{cc}%
0 & M\\
M^{T} & 0
\end{array}
\right)  P^{T},
\]
where $P$ is some permutation matrix and $M$ is a $(0,1)$-matrix. It is clear
that $G$ is bipartite and that the rank of $G$ is twice the rank of $M$. We
say that $G$ is the \emph{bipartite double} of the (possibly directed) graph
with adjacency matrix $M$.

\section{Bipartite graphs with maximal rank}

By Proposition \ref{bou}, a graph state $|G\rangle$ has maximal Schmidt
measure if the bipartite graph $G$ is nonsingular. In the next subsections we
present (in this order) the following bipartite nonsingular graphs:

\begin{itemize}
\item perfect trees;

\item B\i y\i ko\u{g}lu cographs;

\item graphs from anti-Hadamard matrices;

\item certain unicyclic graphs.
\end{itemize}

\subsection{General remarks}

\noindent\textbf{Minimum number of edges. }A \emph{path} in a graph is a
finite sequence of alternating vertices and edges, starting and ending with a
vertex, $v_{1}e_{1}v_{2}e_{2}v_{3}...e_{n-1}v_{n}$, such that every
consecutive pair of vertices $v_{x}$ and $v_{x+1}$ are adjacent and $e_{x}$ is
incident with $v_{x}$ and with $v_{x+1}$. The \emph{length} of a path is the
number of its vertices. A path of length $n$ is denoted by $P_{n}$. A
\emph{cycle} of length $n-1$, that is a path in which $v_{1}=v_{n}$, is
denoted by $C_{n-1}$. A \emph{connected graph} is a graph such that there is a
path between all pairs of vertices. A \emph{connected component} is a maximal
subset of vertices and edges between them that forms a connected graph. On the
base of \cite{cam} (Proposition 18), we can write:

\begin{proposition}
The smallest number of edges of a graph on $n$ vertices associated to a
two-colorable graph state with maximal Schmidt measure is $n/2$ if $n$ is even
and $(n+3)/2$ if $n$ is odd; if the graph is connected then the numbers are
$n-1$ if $n$ is even and $n$ if $n$ is odd. (The graphs in question are obvious.)
\end{proposition}

\noindent\textbf{Random states.} Numerical evidence gives strong support to
the conjecture that the probability that an $n\times n$ $(0,1)$-matrix is
singular is exactly $n^{2}2^{-n}$ (see, \emph{e.g.}, \cite{ko, vo}). Clearly,
$\lim_{n\rightarrow\infty}n^{2}2^{-n}=0$. Based on this conjecture, it is
relatively safe to believe that \emph{almost all two-colorable graph states
have maximal Schmidt measure}.

\noindent\textbf{Nullity preserving operations.} When a vertex and its
incident edges are deleted from a graph $G$, the rank of the resultant graph
cannot exceed $r(G)$ and can decrease by at most $2$. When edges are added to
a graph, the rank of the resultant graph cannot decrease and can increase by
at most $2$ \cite{be}. The following two operations preserve the nullity of a
graph \cite{c}: \emph{(i)} A path of length $6$ is replaced by an edge.
\emph{(ii)} A graph $H$ is a \emph{subgraph} of a graph $G$ if $V(H)\subseteq
V\left(  G\right)  $, $E\left(  H\right)  \subseteq E\left(  G\right)  $ and
every arc in $E\left(  H\right)  $ has both its end-vertices in $V\left(
G\right)  $. A subgraph $H$ of $G$ is an \emph{induced subgraph} if every edge
in $E\left(  G\right)  $, having both vertices in $V\left(  H\right)  $, is
also in $A\left(  H\right)  $. For a graph $G$ having vertex incident with one
edge only, the induced subgraph $H$, obtained by deleting this vertex together
with the vertex adjacent to it, has the same nullity of $G$.

\subsection{Perfect trees}

A \emph{tree} is a graph with no path that starts and ends at the same vertex.
A \emph{matching} is a graph $G$ is a set $S\subseteq E(G)$ no two edges are
incident with a common vertex. A matching $S$ is \emph{perfect} if $|V|=2|S|$.
The \emph{matching number}, denoted by $\beta(G)$, is the largest cardinality
of a matching in $G$. For some classes of graphs (for example, trees) $r(G)$
and $\beta(G)$ are related. A \emph{complete graph} on $n$ vertices, denoted
by $K_{n}$, is a graph such that $E(K_{n})=V(K_{n})\times V(K_{n})$. A
\emph{perfect tree} is defined as follows: the tree $K_{2}$ is perfect; if
$T_{1}$ and $T_{2}$ are perfect trees then the tree obtained by adding an edge
between any vertex of $T_{1}$ and any vertex of $T_{2}$ is also perfect.

The authors of \cite{h} observed that the Schmidt measure of a tree can be
obtained from the size of its smallest vertex cover (that is, the minimum
number of vertices required to cover all edges). Let $T$ be a tree on $n$
vertices. It is well-known that (see, \emph{e.g.}, \cite{b} or \cite{nb},
Theorem 8.1) $r(T)=2\cdot\beta(T)$. Then $r(T)=n$ if and only if $T$ has a
perfect matching. One can show that a tree has a perfect matching if and only
if it is \emph{perfect }(a proof of this result is Lemma 3.2 in \cite{f}).
Given that a graph is bipartite if and only if it has no cycles of odd length,
a tree is bipartite since it is no cycles by definition. The following is a
direct consequence of this reasoning:

\begin{proposition}
\label{p1}Let $T$ be a perfect tree on $n$ vertices. Then $E_{S}(|T\rangle)$
is maximal.
\end{proposition}

\subsection{Cographs}

A graph $G$ is said to be $H$\emph{-free} if $H$ is not an induced subgraph of
$G$. A \emph{cograph} is a $P_{4}$-free graph. An important subclass of
cographs consists of threshold graphs. These are applied to integer
programming, synchronization of parallel processes, \emph{etc.} (see,
\emph{e.g.}, \cite{go}). A connected cograph is bipartite if and only if it is
a complete bipartite graph, and cographs on $n$ vertices and $m$ edges are
recognized in time $O(n+m)$. A simple construction of cographs is given by
Lovasz \cite{l}. For a cograph $G$ on $n$ vertices, Royle \cite{r} and B\i y\i
ko\u{g}lu \cite{bi} settled in the affirmative a conjecture of Sillke
\cite{si}, by proving that if the rows of $M(G)$ are distinct and nonzero then
$r(G)=n$. The cographs with invertible adjacency matrix have been
characterized by B\i y\i ko\u{g}lu (see \cite{bi}, Lemma 4). For this reason,
we refer to these graphs as \emph{B\i y\i ko\u{g}lu cographs}. Defining B\i
y\i ko\u{g}lu cographs is lengthy and it requires a number of notions. The
interested reader is addressed to \cite{bi}.

\begin{proposition}
Let $G$ be the bipartite double of a B\i y\i ko\u{g}lu cograph. Then
$E_{S}(|G\rangle)$ is maximal.
\end{proposition}

\subsection{Graphs from anti-Hadamard matrices}

An \emph{anti-Hadamard matrix} $M$ is an $n\times n$ $(0,1)$-matrix for which
$\mu(M)=\mu(n)$, where $\mu(M)=\sum_{i,j=1}^{n}\left(  M_{i,j}^{-1}\right)
^{2}$ (the square of the Euclidean norm of $M^{-1}$) and $\mu(n)=\max_{M}%
\mu(M)$, with the maximum taken over all invertible $(0,1)$-matrices.
Anti-Hadamard matrices where introduced by Graham and Sloane as $(0,1)$%
-matrices which are \textquotedblleft only just nonsingular\textquotedblright%
\ \cite{sl}. Next is an anti-Hadamard matrix:%
\[
M=\left(
\begin{array}
[c]{cccc}%
1 & 1 & 0 & 1\\
0 & 1 & 1 & 0\\
0 & 0 & 1 & 1\\
1 & 0 & 0 & 1
\end{array}
\right)
\]

\begin{proposition}
Let $G$ be the bipartite double of a digraph whose adjacency matrix is an
anti-Hadamard matrix. Then $E_{S}(|G\rangle)$ is maximal.
\end{proposition}

\subsection{Unicyclic graphs}

A graph is \emph{unicyclic} if it has the same number of vertices and edges. A
unicyclic graph $G$ satisfying one of the two following two properties is said
to be \emph{elementary}: \emph{(i)} The graph $G$ is the cycle $C_{n}$ where
$n\neq0\left(  \operatorname{mod}4\right)  $ \emph{(ii)} The graph $G$ is
constructed as follows: select $t$ vertices from $C_{l}$ such that between two
selected vertices there is an even (possibly $0$) number of vertices, and $t$
is an integer such that $0<t\leq l$ with $l\equiv t\left(  \operatorname{mod}%
2\right)  $; each one of these selected $t$ vertices is then joined to one of
$t$ extra vertices. By a result of \cite{x}, we have the next fact:

\begin{proposition}
If $G$ is an elementary unicyclic graph, or a graph obtained by joining a
vertex of a perfect tree with an arbitrary vertex of an elementary unicyclic
graph, then $E_{S}(|G\rangle)$ is maximal.
\end{proposition}

\section{Graph transformations}

\subsection{Inverse of the adjacency matrix}

\emph{Local complementation} is a graph transformation whose study was
principally carried on by Bouchet \cite{b} and Fon-der-Flaas \cite{fdf}. Local
complementation is important in the context of graph states. Given a graph
$G$, the \emph{neighborhood} of $i\in V(G)$ is $N(i)=\{j:\{i,j\}\in E(G)\}$.
The graph $G_{i}^{c}$ is the \emph{local complement }of $G$ at $i$ if
$V(G_{i}^{c})=V(G)$ and $\{k,l\}\in E(G_{i}^{c})$ if and only if one of the
following two conditions is satisfied: \emph{(i)} $\{k,l\}\in E(G)$ and
$k\notin N(i)$ or $l\notin N(i)$; \emph{(ii)} $\{k,l\}\notin E(G)$ and $k,l\in
N(i)$. The mapping $\gamma_{i}(G)=G_{i}^{c}$ is called \emph{local
complementation} at $i$. Note that $\gamma_{i}(\gamma_{i}(G))=\gamma_{i}%
(G_{i}^{c})=G$. The \emph{Clifford group }$\mathcal{C}_{1}$ on one qubit is
the group of all $2\times2$ unitary matrices $C$, for which $C\sigma
_{u}C^{\dagger}=\pm\sigma_{\pi(u)}$, where $u\in\{x,y,z\}$ and $\pi\in S_{3}$,
the full symmetric group on three symbols. The generators of $\mathcal{C}_{1}$
are the matrices $\tau_{x}=\sqrt{-i\sigma_{x}}$ and $\tau_{z}=\sqrt
{i\sigma_{z}}$. The \emph{Clifford group} $\mathcal{C}_{n}$ on $n$-qubits is
the $n$-fold tensor product of elements of $\mathcal{C}_{1}$. Two graph states
$|G\rangle$ and $|H\rangle$ of dimension $2^{n}$ are said to be
\emph{LC-equivalent} if there is $U\in\mathcal{C}_{n}$ such that $U|G\rangle
U^{\dagger}=|H\rangle$. It may be interesting to mention that the interlace
polynomial is an invariant under local complementation \cite{p}. Hein \emph{et
al.} \cite{h} and Van den Nest \emph{et al.} \cite{vdn} proved the following
link between local complementation and LC-equivalence: two graph states
$|G\rangle$ and $|H\rangle$ are LC-equivalent if and only if $\gamma
_{k}(\gamma_{j}(\ldots(\gamma_{i}(G)))=H$, where $\gamma_{k},\gamma_{j}%
\ldots\gamma_{i}$ is a sequence of local complementations at vertices
$k,j,...,i\in V(G)$. Let $G^{I}$ be the graph on $n$ vertices whose adjacency
matrix is obtained from $M(G)^{-1}$ by changing the sign of the negative
entries. Jaeger proved that if $r(G)=n$ then $r(G^{I})=n$ and $G^{I}$ can be
obtained from $G$ by a sequence of local complementations \cite{j} (see also
\cite{bo}). This implies the following:

\begin{proposition}
\label{inv}Let $|G\rangle$ be a two-colorable graph state with maximal Schmidt
measure. Then $|G^{I}\rangle$ has maximal Schmidt measure. Moreover,
$|G\rangle$ and $|G^{I}\rangle$ are LC-equivalent.
\end{proposition}

\subsection{Line graph}

The \emph{line graph }of a graph $G$, denoted by $L(G)$, is the graph whose
set of vertices is $E(G)$ and $\{\{i,j\},\{k,l\}\}\in E(L(G))$ if and only if
one of the following conditions is satisfied:$\ j=k$, $j=l$, $i=k$ or $i=l$.
The rank of the line graph of a graph on $n$ vertices is at least $n-2$
(\cite{cam}, Proposition 14). A \emph{clique} is an induced complete subgraph.
In a graph $G$, a vertex $i$ is a \emph{cutpoint} if the graph $G\backslash
i$, obtained by deleting $i$ and all edges incident with $i$, has more
connected components than $G$. Given a graph $G$, we have $G=L(T)$ for some
tree $T$ if and only if $E(G)$ can be partitioned into a set of cliques with
the property that any vertex is in either one or two cliques; if a vertex is
in two cliques then it is a cutpoint. Gutman and Sciriha proved that if $T$ is
a tree then $L(T)$ is either nonsingular or it has nullity $1$ (\cite{gs},
Theorem 2.1). However, the line graph transformation applied to trees it is
not directly relevant to our context, since we need $L(T)$ to be bipartite. In
fact, it is clear that $L(T)$ is bipartite (and a path) if and only if $T$
itself is a path.

\subsection{Switching}

Proposition \ref{inv} gives a method to obtain a graph state (not necessarily
two-colorable) with maximal Schmidt measure from a two-colorable graph state
with the same entanglement. The graph operation considered in Proposition
\ref{inv} is then guaranteed to preserve the amount of entanglement (which is
that case is maximal). Is there a simple operation that is guaranteed to
change the amount of entanglement? \emph{Switching} is a graph operation
introduced by Van Lint and Seidel \cite{vls} (see \cite{ha}, for a survey).
The graph $G_{i}^{s}$ is the \emph{switching} of $G$ at $i$ if $V(G_{i}%
^{s})=V(G)$ and $\{k,l\}\in E(G_{i}^{s})$ if and only if one of the following
two conditions is satisfied: \emph{(i)} $\{k,l\}\in E(G)$ and, $k\neq j$ and
$l\neq i$; \emph{(ii)} $\{k,l\}\notin E(G)$ and, $k=i$ or $l=i$. The mapping
$s_{i}(G)=G_{i}^{s}$ is called \emph{switching} at $i$. It follows directly
from the definition that $A(G_{i}^{c})=A(G)-D+C$, where $D=%
{\textstyle\sum\nolimits_{j\in N(i)}}
e_{i}e_{j}^{T}+\left(
{\textstyle\sum\nolimits_{j\in N(i)}}
e_{i}e_{j}^{T}\right)  ^{T}$and $C=%
{\textstyle\sum\nolimits_{j\notin N(i)}}
e_{i}e_{j}^{T}+\left(
{\textstyle\sum\nolimits_{j\notin N(i)}}
e_{i}e_{j}^{T}\right)  ^{T}$. Deciding if a graph $G$ can be obtained from a
graph $H$ by switching is polynomial time equivalent to graph isomorphism
\cite{cc}. The \emph{switching operator} with respect to $k$ is the unitary
operator $T_{k}:\mathbb{C}^{2^{n}}\longrightarrow\mathbb{C}^{2^{n}}$ defined
as
\[
T_{k}|x_{1}x_{2}...x_{n}\rangle=\left(  -1\right)  ^{x_{k}\sum_{i=1:i\neq
k}^{n}x_{i}}|x_{1}x_{2}...x_{n}\rangle,
\]
for $x_{1},x_{2},...,x_{n}\in\{0,1\}$. It is easy to see that for the graph
states $|G\rangle$ and $|G_{k}^{s}\rangle$, we have $T_{k}|G\rangle=|G_{k}%
^{s}\rangle$. A group $\mathcal{G}$ acting (on the left) on a set $\Omega$ is
\emph{transitive} if for every $\alpha,\beta\in\Omega$ there is $g\in
\mathcal{G}$ such that $g\alpha=\beta$. Let $\Omega_{n}%
=\{G:V(G)=\{1,2,...,n\}\}$. Ehrenfeucht \emph{et al.} \cite{eh} prove that the
composition of local complementation and switching forms a transitive group
acting on the set $\Omega_{n}$. Let us denote by $\Omega_{k}^{S}$ the set of
graph states of dimension $k$. We can then write:

\begin{proposition}
\label{swi}The composition of elements from the local Clifford group and
switching operators forms a transitive group acting on the set $\Omega_{k}%
^{S}$.
\end{proposition}

The meaning of this observation is clear:\ from a graph state $|G\rangle$ of a
given dimension one can obtain any other graph state $|H\rangle$ of the same
dimension by the application of elements from the local Clifford group and
switching operators. The \emph{signed adjacency matrix} of a graph $G$ is the
matrix $A^{+}(G)$ such that $A^{+}(G)_{i,j}=1$ if $\{i,j\}\in E(G)$,
$A^{+}(G)_{i,j}=-1$ if $\{i,j\}\notin E(G)$. The spectrum of $A^{+}(G)$ is
equal to the spectrum of $A^{+}(H)$ if the graphs $G$ and $H$ are obtained one
from the other one by switching. As a consequence, the spectrum of $A^{+}(G)$
does not seem to contain much information about the entanglement properties of
$|G\rangle$. In the figure below are drawn all (nonisomorphic) graphs on four
vertices. An arrow between two graphs indicates that the graphs are in the
same switching class; a dotted arrow indicates that the graphs are in the same
local complementation class:%
\begin{center}
\includegraphics[
height=4.2301cm,
width=6.1997cm
]%
{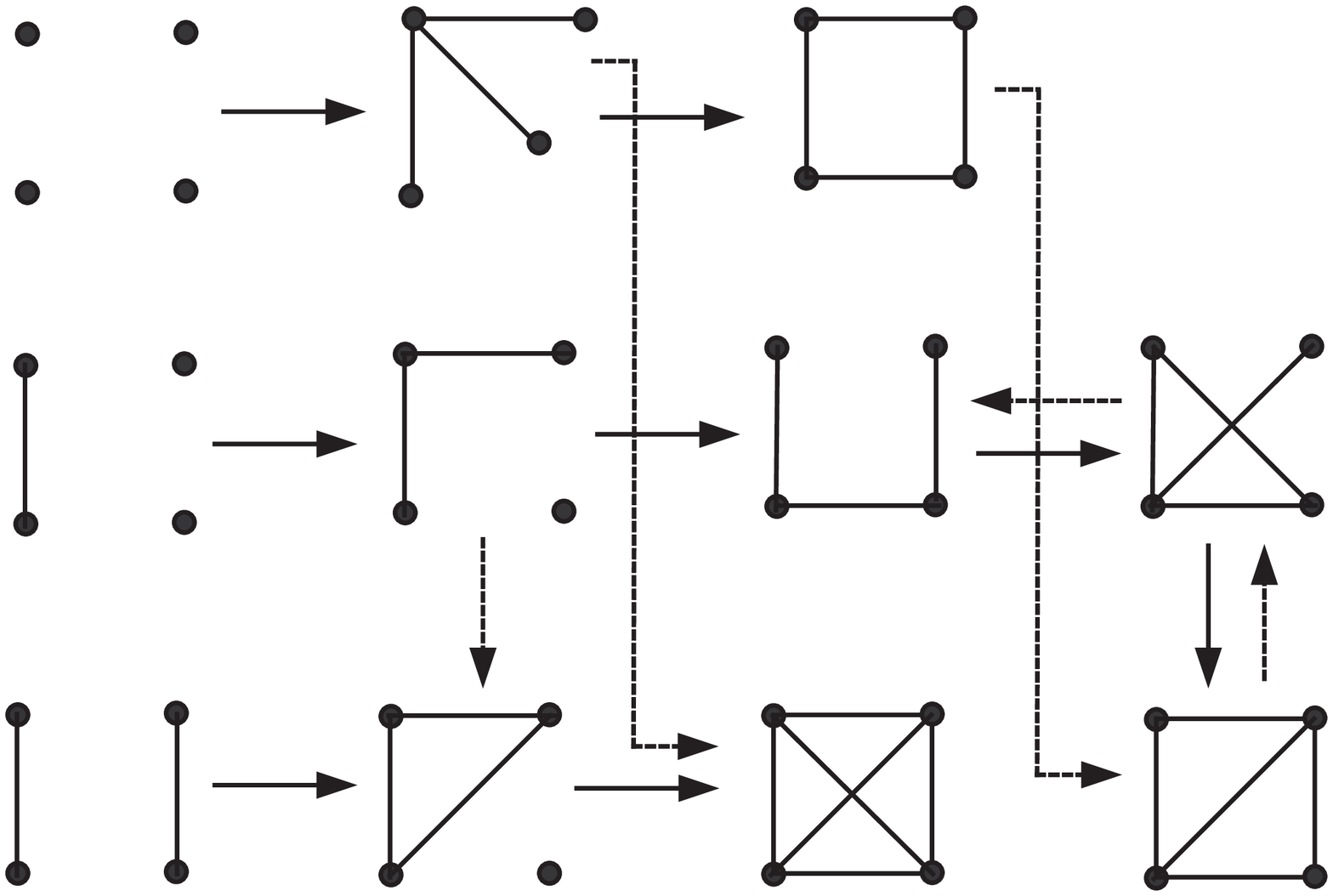}%
\end{center}
Notice that there are some graphs which are linked by both, local
complementation and switching. The \emph{chromatic number} of a graph $G$,
denoted by $\chi(G)$, is the minimum number of colors needed to color the
vertices of $G$ such that adjacent vertices have different colors. Let $[G]$
be the \emph{switching class} of $G$, that is the set of all graphs obtained
by a sequence of switching on a graph $G$. If $\chi(G)=k$ then $2\leq
\chi(H)\leq2k$, for all $H\in\lbrack G]$. If a switching class has a graph
with chromatic number larger than $4$ then is does not contain a bipartite
graph, but the converse it is not necessarily true (\cite{ha}, Lemma 3.30). It
is not immediate to understand how entanglement is modified by switching.
However, on the light of Proposition \ref{swi}, switching is potentially
useful in classifying graph states (see \cite{d}, for a work on the
classification of graph states).

\begin{acknowledgments}
The author would like to thank Roland Bacher, Jens Eisert and Matthew Joffrey
Parker for helpful comments, and Sibasish Ghosh for introducing him to graph states.
\end{acknowledgments}

\end{document}